\documentclass[ ,showpacs
 ,secnumarabic, nobibnotes, nofootinbib, aps, prd]{revtex4}
\usepackage{bm}
\usepackage{graphicx}

\begin{document}

\title{Parton distribution functions and quark orbital motion
\footnote{Paper is accepted for publication in Eur.Phys.J. C}}
\author{Petr Z\'{a}vada}
\email{zavada@fzu.cz}
\affiliation{Institute of Physics, Academy of Sciences of the Czech Republic, \\
Na Slovance 2, CZ-182 21 Prague 8}

\begin{abstract}
Covariant version of the quark-parton model is studied. Dependence of the
structure functions and parton distributions on the 3D quark intrinsic
motion is discussed. The important role of the quark orbital momentum, which
is a particular case of intrinsic motion, appears as a direct consequence of
the covariant description. Effect of orbital motion is substantial
especially for polarized structure functions. At the same time, the
procedure for obtaining the quark momentum distributions of polarized quarks
from the combination of polarized and unpolarized structure functions is
suggested.
\end{abstract}

\maketitle


\section{Introduction}

The nucleon structure functions are basic tool for understanding the nucleon
internal structure in the language of QCD. And at the same time, the
measuring and analysis of the structure functions represent the important
experimental test of this theory. Unpolarized nucleon structure functions
are known with high accuracy in very broad kinematical region, but in recent
years also some precision measurements on the polarized structure functions
have been completed \cite{e142,e154,her1,her2,adeva,e143,E155}. For present
status of the nucleon spin structure see e.g. \cite{spin04} and citations
therein. The more formal aspects of the nucleon structure functions are
explained in \cite{efrem}. In fact only the complete set of the four
electromagnetic unpolarized and polarized structure functions $%
F_{1},F_{2},g_{1}$ and $g_{2}$ can give a consistent picture of the nucleon.
However, this picture is usually drawn in terms of the distribution
functions, which are connected with the structure functions by some
model-dependent way. Distribution functions are not directly accessible from
the experiment and model, which is normally applied for their extraction
from the structure functions is the well known quark-parton model (QPM).
Application of this model for analysis and interpretation of the unpolarized
data does not create any contradiction. On the other hand, the situation is
much less clear in the case of spin functions $g_{1}$ and $g_{2}$.

In our previous study \cite{zav1,zav2} we have suggested, that a reasonable
explanation of the experimentally measured spin functions $g_{1},$ $g_{2}$
is possible in terms of a generalized covariant QPM, in which the quark
intrinsic motion (i.e. 3D motion with respect to the nucleon rest frame) is
consistently taken into account. Therefore the quark transversal momentum
appears in this approach on the same level as the longitudinal one. The
quarks are represented by the free Dirac spinors, which allows to obtain
exact and covariant solution for relations between the quark momentum
distribution functions and the structure functions accessible from
experiment. In this way the model (in its present LO version) contains no
dynamics but only \textquotedblleft exact\textquotedblright\ kinematics of
quarks, so it can be effective tool for analysis and interpretation of the
experimental data on structure functions, particularly for separating
effects of the dynamics (QCD) from effects of the kinematics. This point of
view is well supported by our previous results:

\textit{a)} In the cited papers we showed, that the model simply implies the
well known sum rules (Wanzura-Wilczek, Efremov-Leader-Teryaev,
Burkhardt-Cottingham) for the spin functions $g_{1},$ $g_{2}$.

\textit{b}) Simultaneously, we showed that the same set of assumptions
implies rather substantial dependence of the first moment $\Gamma _{1}$ of
the function $g_{1}$ on the kinematical effects.

\textit{c)} Further, we showed that the model allows to calculate the
functions $g_{1},$ $g_{2}$ from the unpolarized valence quark distributions
and the result is quite compatible with the experimental data.

\textit{d)} In the paper \cite{tra}\ we showed that the model allows to
relate the transversity distribution to some other structure functions.

These results cannot be obtained from the standard versions of the QPM
(naive or the QCD improved), which are currently used for the analysis of
experimental data on structure functions. The reason is, that the standard
QPM is based on the simplified and non-covariant kinematics in the infinite
momentum frame (IMF), which does not allow to properly take into account the
quark intrinsic or orbital motion.

The subject of our previous study was the question: \textit{What is the
dependence of the structure functions on quark intrinsic motion?} The aim of
the present paper is a discussion of related problems:

\textit{1. How to extract information about the quark intrinsic motion from
the experimentally measured structure functions?}

\textit{2. What is the role of the quark orbital momentum, which is a
particular case of intrinsic motion?}

The paper is organized as follows. In the first part of Sec. \ref{sec2} the
basic formulas, which follow from the generalized QPM, are presented.
Resulting general\ covariant relations are compared with their limiting
case, which is represented by the standard formulation of the QPM in the
IMF. In the next part of the section the relations for calculation of 3D
quark momentum distributions from the structure functions are derived. The
quark momentum distributions of positively and negatively polarized quarks
are separately obtained from the combination of structure functions $%
F_{2},g_{1}$ or corresponding parton distributions $q(x),\Delta q(x)$. The
particular form of the quark intrinsic motion is the orbital momentum. In
Sec. \ref{sec3} the role of the quark orbital momentum in covariant
description is discussed and it is shown, why its contribution to the total
quark angular momentum can be quite substantial. It is demonstrated, that
the orbital motion is an inseparable part of the covariant approach. The
last section is devoted to a short summary and conclusion. In fact, this
paper is inspired by many previous papers, see e.g. \cite%
{sehgal,ratc,abbas,kep,casu,bqma,brod,waka,waka1,song,ji}, in which problem
of quark orbital momentum in the context of nucleon spin was recognized and
studied.

\section{Structure functions and intrinsic quark motion}

\label{sec2}In our previous study \cite{zav0,zav1,zav2} of the proton
structure functions we showed, how these functions depend on the intrinsic
motion of quarks. The quarks in the suggested model are represented by the
free fermions, which are in the nucleon rest frame described by the set of
distribution functions with spheric symmetry $G_{k}^{\pm }(p_{0})d^{3}p$,
where $p_{0}=\sqrt{m^{2}+\mathbf{p}^{2}}$ and symbol $k$ represents the
quark and antiquark flavors. These distributions measure the probability to
find a quark of given flavor in the state%
\begin{equation}
u\left( \mathbf{p},\lambda \mathbf{n}\right) =\frac{1}{\sqrt{N}}\left( 
\begin{array}{c}
\phi _{\lambda \mathbf{n}} \\ 
\frac{\mathbf{p}\mathbf{\sigma }}{p_{0}+m}\phi _{\lambda \mathbf{n}}%
\end{array}%
\right) ;\qquad \frac{1}{2}\mathbf{n\sigma }\phi _{\lambda \mathbf{n}%
}=\lambda \phi _{\lambda \mathbf{n}},\qquad N=\frac{2p_{0}}{p_{0}+m},
\label{sp8}
\end{equation}%
where $m$ and $p$\ are the quark mass and momentum, $\lambda =\pm 1/2,$ $%
\phi _{\lambda \mathbf{n}}^{\dagger }\phi _{\lambda \mathbf{n}}=1$ and$\ 
\mathbf{n}$ coincides with the direction of nucleon polarization. The
distributions with the corresponding quark (and antiquark) charges $e_{k}$
allow to define the generic functions $G$ and $\Delta G$\footnote{%
In the papers \cite{zav1,zav2} we used different notation for the
distributions defined by Eqs.(\ref{sp8a}) and (\ref{sp9}): $G_{k}^{\pm
},\Delta G_{k}$ and $\Delta G$ were denoted as $h_{k\pm },\Delta h_{k}$ and $%
H$. Apart of that we assumed for simplicity that only three (valence) quarks
contribute to the sums (\ref{sp8a}) and (\ref{sp9}). In present paper we
assume contribution of all the quarks and antiquarks, but apparently general
form of the relations like (\ref{cra31}) - (\ref{cr31}) is in the LO
approach independent of chosen set of quarks.},%
\begin{equation}
G(p_{0})=\sum_{k}e_{k}^{2}G_{k}(p_{0}),\quad G_{k}(p_{0})\equiv
G_{k}^{+}(p_{0})+G_{k}^{-}(p_{0}),  \label{sp8a}
\end{equation}%
\begin{equation}
\Delta G(p_{0})=\sum_{k}e_{k}^{2}\Delta G_{k}(p_{0}),\quad \Delta
G_{k}(p_{0})\equiv G_{k}^{+}(p_{0})-G_{k}^{-}(p_{0}),  \label{sp9}
\end{equation}%
from which the structure functions can be obtained. If $q$ is momentum of
the photon absorbed by the nucleon of the momentum $P$ and mass $M,$ in
which the phase space of quarks is controlled by the distributions $%
G_{k}^{\pm }(p_{0})d^{3}p$, then there are the following representations of
corresponding LO structure functions.

\textit{A. Manifestly covariant representation}

\textit{i) unpolarized structure functions:} 
\begin{equation}
F_{1}(x)=\frac{M}{2}\left( A+\frac{B}{\gamma }\right) ,\qquad F_{2}(x)=\frac{%
Pq}{2M\gamma }\left( A+\frac{3B}{\gamma }\right) ,  \label{sp9b}
\end{equation}%
where%
\begin{equation}
A=\frac{1}{Pq}\int G\left( \frac{Pp}{M}\right) \left[ pq-m^{2}\right] \delta
\left( \frac{pq}{Pq}-x\right) \frac{d^{3}p}{p_{0}},  \label{sp9c}
\end{equation}%
\begin{equation}
B=\frac{1}{Pq}\int G\left( \frac{pP}{M}\right) \left[ \left( \frac{Pp}{M}%
\right) ^{2}+\frac{\left( Pp\right) \left( Pq\right) }{M^{2}}-\frac{pq}{2}%
\right] \delta \left( \frac{pq}{Pq}-x\right) \frac{d^{3}p}{p_{0}}
\label{sp9d}
\end{equation}%
and%
\begin{equation}
\gamma =1-\left( \frac{Pq}{Mq}\right) ^{2}.  \label{sp9e}
\end{equation}%
The functions $F_{1}=MW_{1}$ and $F_{2}=\left( Pq/M\right) W_{2}$ follow
from the tensor equation%
\begin{eqnarray}
&&\left( -g_{\alpha \beta }+\frac{q_{\alpha }q_{\beta }}{q^{2}}\right)
W_{1}+\left( P_{\alpha }-\frac{Pq}{q^{2}}q_{\alpha }\right) \left( P_{\beta
}-\frac{Pq}{q^{2}}q_{\beta }\right) \frac{W_{2}}{M^{2}}  \label{sp9f} \\
&=&\int G\left( \frac{pP}{M}\right) \left[ 2p_{\alpha }p_{\beta }+p_{\alpha
}q_{\beta }+q_{\alpha }p_{\beta }-g_{\alpha \beta }pq\right] \delta \left(
\left( p+q\right) ^{2}-m^{2}\right) \frac{d^{3}p}{p_{0}}.  \nonumber
\end{eqnarray}%
By modification of the delta function term%
\begin{equation}
\delta \left( \left( p+q\right) ^{2}-m^{2}\right) =\delta \left(
2pq+q^{2}\right) =\delta \left( 2Pq\left( \frac{pq}{Pq}-\frac{Q^{2}}{2Pq}%
\right) \right) =\frac{1}{2Pq}\delta \left( \frac{pq}{Pq}-x\right) ;\qquad
q^{2}=-Q^{2},\quad x=\frac{Q^{2}}{2Pq},  \label{sp9g}
\end{equation}%
the dependence on the Bjorken $x$ is introduced. Then contracting with the
tensors $g_{\alpha \beta }$ and $P_{\alpha }P_{\beta }$ gives the set of two
equations, which determine the functions $F_{1},F_{2}$ in accordance with
Eqs. (\ref{sp9b})-(\ref{sp9e}).

\textit{ii) polarized structure functions:}

As follows from \cite{zav1} the corresponding spin functions in covariant
form read%
\begin{equation}
g_{1}=Pq\left( G_{S}-\frac{Pq}{qS}G_{P}\right) ,\qquad g_{2}=\frac{\left(
Pq\right) ^{2}}{qS}G_{P},  \label{cra31}
\end{equation}%
where $S$ is the nucleon spin polarization vector and the functions $%
G_{P},G_{S}$\ are defined as%
\begin{equation}
G_{P}=\frac{m}{2Pq}\int \Delta G\left( \frac{pP}{M}\right) \frac{pS}{pP+mM}%
\left[ 1+\frac{1}{mM}\left( pP-\frac{pu}{qu}Pq\right) \right] \delta \left( 
\frac{pq}{Pq}-x\right) \frac{d^{3}p}{p_{0}},  \label{cr30}
\end{equation}%
\begin{equation}
G_{S}=\frac{m}{2Pq}\int \Delta G\left( \frac{pP}{M}\right) \left[ 1+\frac{pS%
}{pP+mM}\frac{M}{m}\left( pS-\frac{pu}{qu}qS\right) \right] \delta \left( 
\frac{pq}{Pq}-x\right) \frac{d^{3}p}{p_{0}};  \label{cr31}
\end{equation}%
\[
u=q+\left( qS\right) S-\frac{\left( Pq\right) }{M^{2}}P. 
\]

\textit{B. Rest frame representation for} $Q^{2}\gg 4M^{2}x^{2}$

As follows from the Appendix in \cite{zav1}, if $Q^{2}\gg 4M^{2}x^{2}$ and
the above integrals are expressed in terms of the nucleon rest frame
variables, then one can substitute 
\[
\frac{pq}{Pq}\rightarrow \frac{p_{0}+p_{1}}{M} 
\]%
and the structure functions are simplified as:%
\begin{eqnarray}
F_{1}(x) &=&\frac{Mx}{2}\int G(p_{0})\delta \left( \frac{p_{0}+p_{1}}{M}%
-x\right) \frac{d^{3}p}{p_{0}},  \label{sp9h} \\
F_{2}(x) &=&Mx^{2}\int G(p_{0})\delta \left( \frac{p_{0}+p_{1}}{M}-x\right) 
\frac{d^{3}p}{p_{0}},  \label{sp9a} \\
g_{1}(x) &=&\frac{1}{2}\int \Delta G(p_{0})\left( m+p_{1}+\frac{p_{1}^{2}}{%
p_{0}+m}\right) \delta \left( \frac{p_{0}+p_{1}}{M}-x\right) \frac{d^{3}p}{%
p_{0}},  \label{sp10} \\
g_{2}(x) &=&-\frac{1}{2}\int \Delta G(p_{0})\left( p_{1}+\frac{%
p_{1}^{2}-p_{T}^{2}/2}{p_{0}+m}\right) \delta \left( \frac{p_{0}+p_{1}}{M}%
-x\right) \frac{d^{3}p}{p_{0}},  \label{sp11}
\end{eqnarray}%
where the $p_{1}$ and $p_{T}$ are longitudinal and transversal quark
momentum components. These structure functions consist of terms like%
\begin{eqnarray}
q(x) &=&Mx\int G_{q}(p_{0})\delta \left( \frac{p_{0}+p_{1}}{M}-x\right) 
\frac{d^{3}p}{p_{0}},  \label{sp11e} \\
\Delta q(x) &=&\int \Delta G_{q}(p_{0})\left( m+p_{1}+\frac{p_{1}^{2}}{%
p_{0}+m}\right) \delta \left( \frac{p_{0}+p_{1}}{M}-x\right) \frac{d^{3}p}{%
p_{0}},  \label{sp11d}
\end{eqnarray}%
which correspond to the contributions from different quark flavors, $q=u,%
\bar{u},d,\bar{d},s,\bar{s},...$ Let us remark, in limit of the IMF approach
(see next paragraph) these functions represent probabilistic distributions
of the quark momentum in terms of the fraction $x$ of the nucleon momentum, $%
p=xP$. However content and interpretation of the functions (\ref{sp11e}),(%
\ref{sp11d}) depending on the Bjorken $x$ is more complex; their form
reflects in a non-trivial way intrinsic 3D motion of quarks.

\textit{C. Standard IMF representation}

The usual formulation of the QPM gives the known relations between the
structure functions and the parton distribution functions \cite{efrem}:%
\begin{equation}
F_{1}(x)=\frac{1}{2}\sum_{q}e_{q}^{2}q(x),\qquad
F_{2}(x)=x\sum_{q}e_{q}^{2}q(x),  \label{sp11a}
\end{equation}%
\begin{equation}
g_{1}(x)=\frac{1}{2}\sum_{q}e_{q}^{2}\Delta q(x),\qquad g_{2}(x)=0,
\label{sp11b}
\end{equation}%
where the functions 
\begin{equation}
q(x)=q^{+}(x)+q^{-}(x),\qquad \Delta q(x)=q^{+}(x)-q^{-}(x)  \label{sp11c}
\end{equation}%
represent probabilistic distributions of the of the quark momentum fraction $%
x$ in the IMF. In the Appendix \ref{app0} we have proved that these
relations represent the particular, limiting case of the covariant relations
(\ref{sp9b}) and (\ref{cra31}).

The three versions of the relations between the structure functions and the
quark distributions can be compared:

\textit{a)} If we skip the function $g_{2}$ in the version \textit{C}, then
the relations (\ref{sp11a}) and (\ref{sp11b}) practically represent identity
between the structure functions and distributions of the quark momentum
fraction. Such simple relations are valid only for the IMF approach based on
the approximation (\ref{b2}), which means that the quark intrinsic motion is
suppressed. In more general versions \textit{A} and \textit{B}, where the
intrinsic motion is allowed, the relations are more complex. The intrinsic
motion strongly modifies also the $g_{2}$. In the version \textit{C} there
is $g_{2}(x)=0$, but $g_{2}(x)\neq 0$ in the \textit{A} and \textit{B}.

\textit{b)} The version \textit{B} allows to easily calculate the
(substantial) dependence of the first moment $\Gamma _{1}$ on the rate of
intrinsic motion. A more detailed discussion follows in the next section.
The same approach implies that functions $g_{1}$ and $g_{2}$ for massless
quarks satisfy the relation equivalent to the Wanzura-Wilczek term and obey
some well known sum rules, that is shown in \cite{zav1}.

\textit{c)} The functions $F_{1}$ and $F_{2}$ exactly satisfy the
Callan-Gross relation $F_{2}(x)/F_{1}(x)=2x$ in the versions \textit{B} and
C, but this relation is satisfied only approximately in the \textit{A}: $%
F_{2}(x)/F_{1}(x)\approx 2x+O\left( 4M^{2}x^{2}/Q^{2}\right) $.

The task which was solved in different approximations above can be
formulated: How to obtain the structure functions $F_{1},F_{2}$ and $%
g_{1},g_{2}$ from the probabilistic distributions $G$ and $\Delta G$ defined
by Eqs. (\ref{sp8a}) and (\ref{sp9})? In the next we will study the inverse
problem, the aim is to find out a rule for obtaining the distribution
functions $G$ and $\Delta G$ from the structure functions. In the present
paper we consider the functions $F_{2}$ and $g_{1}$ represented by Eqs. (\ref%
{sp9a}) and (\ref{sp10}). As follows from the Appendix A in \cite{zav2}, the
function%
\begin{equation}
V_{n}(x)=\int K(p_{0})\left( \frac{p_{0}}{M}\right) ^{n}\delta \left( \frac{%
p_{0}+p_{1}}{M}-x\right) d^{3}p  \label{gp1}
\end{equation}%
satisfies%
\begin{equation}
V_{n}^{\prime }(x_{\pm })x_{\pm }=\mp 2\pi MK(\xi )\xi \sqrt{\xi ^{2}-m^{2}}%
\left( \frac{\xi }{M}\right) ^{n};\qquad x_{\pm }=\frac{\xi \pm \sqrt{\xi
^{2}-m^{2}}}{M}.  \label{gp2}
\end{equation}%
In this section we consider only the case $m\rightarrow 0$, then%
\begin{equation}
V_{n}^{\prime }(x)x=-2\pi MK(\xi )\xi ^{2}\left( \frac{\xi }{M}\right)
^{n};\qquad x=\frac{2\xi }{M}.  \label{gp3}
\end{equation}%
As we shall see below, with the use of this relation one can obtain the
probabilistic distributions $G(p)$ and $\Delta G(p)$ from the experimentally
measured structure functions. The same procedure will be applied to get the $%
G_{q}(p)$ and $\Delta G_{q}(p)$ from the usual parton distributions $q(x)$
and $\Delta q(x)$ defined by Eqs. (\ref{sp11a}) - (\ref{sp11b}).

Let us remark that in present stage the QCD evolution is not included into
the model. However, this fact does not represent any restriction for the
present purpose - to obtain information about distributions of quarks at
some scale $Q^{2}$ from the structure functions measured at the same $Q^{2}$%
. Distribution of the gluons is another part of the nucleon picture. But
since our present discussion is directed to the relation between the
structure functions and corresponding (quark) distributions at given scale,
the gluon distribution is left aside.

\subsection{Momentum distribution from structure function $F_{2}$}

In an accordance with the definition (\ref{gp1}) in which the distribution $%
K(p_{0})$\ is substituted\ by the $G(p_{0})$, the structure function (\ref%
{sp9a}) can be written in the form

\begin{equation}
F_{2}(x)=x^{2}V_{-1}(x).  \label{gp5}
\end{equation}%
Then, with the use of the relation (\ref{gp3}) one gets%
\begin{equation}
G(p)=-\frac{1}{\pi M^{3}}\left( \frac{F_{2}(x)}{x^{2}}\right) ^{\prime }=%
\frac{1}{\pi M^{3}x^{2}}\left( \frac{2F_{2}(x)}{x}-F_{2}^{\prime }(x)\right)
;\qquad x=\frac{2p}{M},\qquad p\equiv \sqrt{\mathbf{p}^{2}}=p_{0},
\label{gp8}
\end{equation}%
which in terms of the quark distributions means%
\begin{equation}
G_{q}(p)=-\frac{1}{\pi M^{3}}\left( \frac{q(x)}{x}\right) ^{\prime }=\frac{1%
}{\pi M^{3}x^{2}}\left( q(x)-xq^{\prime }(x)\right) .  \label{gp9}
\end{equation}%
Probability distribution $G_{q}$ measures number of quarks of flavor $q$ in
the element $d^{3}p$. Since $d^{3}p=4\pi p^{2}dp$, the distribution
measuring the number of quarks in the element $dp/M$ reads:%
\begin{equation}
P_{q}(p)=4\pi p^{2}MG_{q}(p)=-x^{2}\left( \frac{q(x)}{x}\right) ^{\prime
}=q(x)-xq^{\prime }(x).  \label{gp9a}
\end{equation}%
The probability distribution $G_{q}(p)$ is positive, so the last relation
implies%
\begin{equation}
\left( \frac{q(x)}{x}\right) ^{\prime }\leq 0.  \label{gp9b}
\end{equation}%
Let us note, the maximum value of quark momentum is $p_{\max }=M/2$, which
is a consequence of the kinematics in the nucleon rest frame, where the
single quark momentum must be compensated by the momentum of the other
partons.

\begin{figure}[tbp]
\includegraphics[width=18cm]{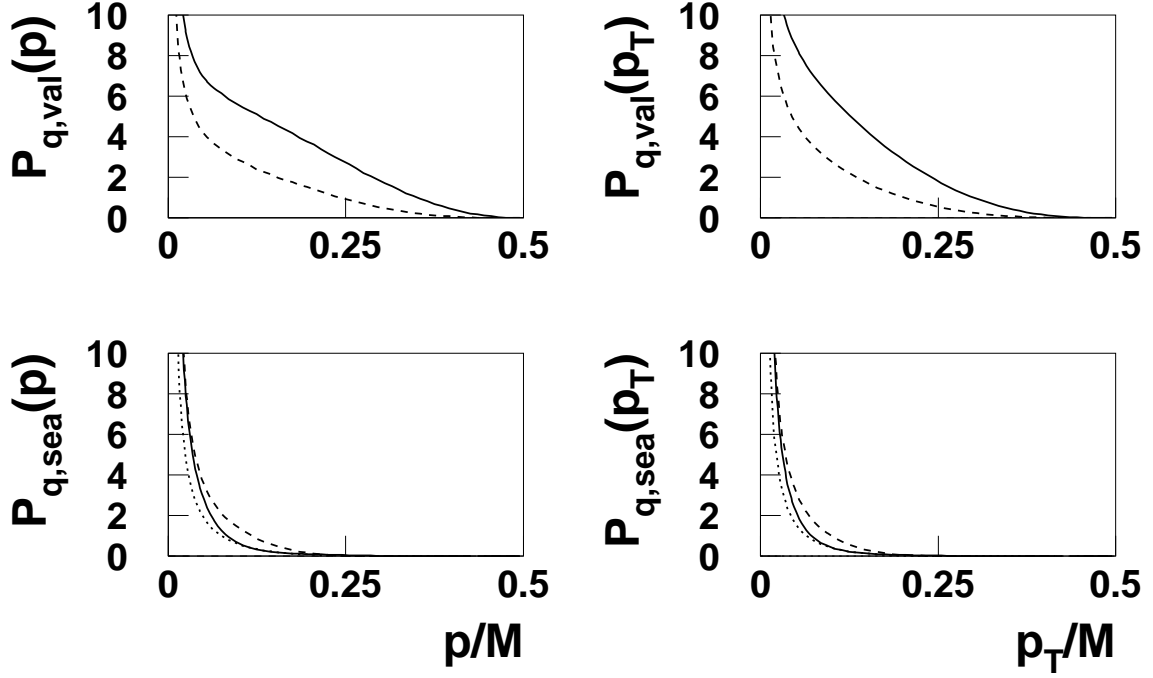}
\caption{The quark momentum distributions in the proton rest frame: the $p$
and $p_{T}$ distributions for valence quarks $P_{q,val}=P_{q}-P_{\bar{q}}$
and sea quarks $P_{\bar{q}}$ at $Q^{2}=4GeV^{2}$. Notation: $u,\bar{u}$\ -
solid line, $d,\bar{d}$ - dashed line, $\bar{s}$ - dotted line.}
\label{fgr1}
\end{figure}

Another quantity, which can be obtained, is the distribution of the quark
transversal momentum. Obviously the integral%
\begin{equation}
\frac{dN_{q}}{dp_{T}^{2}}=\int G_{q}(p)\delta \left(
p_{2}^{2}+p_{3}^{2}-p_{T}^{2}\right) d^{3}p,  \label{gp12}
\end{equation}%
which represents the number of quarks in the element $dp_{T}^{2}$, can be
modified as%
\begin{equation}
\frac{dN_{q}}{dp_{T}^{2}}=2\pi \int_{0}^{\sqrt{p_{\max }^{2}-p_{T}^{2}}%
}G_{q}\left( \sqrt{p_{1}^{2}+p_{T}^{2}}\right) dp_{1}.  \label{gp13}
\end{equation}%
It follows, that the distribution corresponding to the number of quarks in
the element $dp_{T}/M$ reads:%
\begin{equation}
P_{q}(p_{T})=M\frac{dN_{q}}{dp_{T}}=4\pi p_{T}M\int_{0}^{\sqrt{p_{\max
}^{2}-p_{T}^{2}}}G_{q}\left( \sqrt{p_{1}^{2}+p_{T}^{2}}\right) dp_{1}.
\label{gp14}
\end{equation}%
Then with the use of Eq. (\ref{gp9a}) one gets the distribution%
\begin{equation}
P_{q}(p_{T})=\frac{4p_{T}}{M^{2}}\int_{0}^{\sqrt{p_{\max }^{2}-p_{T}^{2}}}%
\frac{1}{x^{2}}\left( q(x)-xq^{\prime }(x)\right) dp_{1};\qquad x=\frac{2%
\sqrt{p_{1}^{2}+p_{T}^{2}}}{M}.  \label{gp16}
\end{equation}%
In Fig. \ref{fgr1} \ the distributions (\ref{gp9a}) and (\ref{gp16}) are
displayed for the valence and sea quarks. \ As an input we used the standard
parameterization \cite{pdf} of the parton distribution functions $q(x),$ $%
\bar{q}(x)$ (LO at the scale $4GeV^{2}$). Resulting distributions $P_{q},P_{%
\bar{q}}$ are positive, it means that the input distributions $q,$ $\bar{q}$
satisfy the constraint (\ref{gp9b}).

Using the Eq. (\ref{gp9}) one can calculate the mean values%
\begin{equation}
\left\langle p\right\rangle _{q}=\frac{\int pG_{q}(p)d^{3}p}{\int
G_{q}(p)d^{3}p}=\frac{M}{2}\frac{\int_{0}^{1}x\left( q(x)-xq^{\prime
}(x)\right) dx}{\int_{0}^{1}\left( q(x)-xq^{\prime }(x)\right) dx}.
\label{gp16a}
\end{equation}%
In the case of sea quarks extrapolation of the distribution functions for $%
x\rightarrow 0$ gives a divergent integral in the denominator, it follows
that $\left\langle p\right\rangle _{sea}\rightarrow 0$. For the valence
quarks $q_{val}=q-$ $\bar{q}$ this integral converges and integration by
parts gives%
\begin{equation}
\left\langle p\right\rangle _{q,val}=\frac{3M}{4}\frac{%
\int_{0}^{1}xq_{val}(x)dx}{\int_{0}^{1}q_{val}(x)dx}.  \label{gp16b}
\end{equation}%
Calculation $\left\langle p\right\rangle _{q,val}$ gives roughly $0.11GeV/c$
for $u$ and $0.083GeV/c$ for $d$ quarks. Since $G_{q}(p)$ has rotational
symmetry, average transversal momentum can be calculated as $\left\langle
p_{T}\right\rangle =\pi /4\cdot \left\langle p\right\rangle $.

\subsection{Momentum distribution from structure function $g_{1}$}

\label{sec22}In the paper \cite{zav2}, Eq. (44), we proved that%
\begin{equation}
g_{1}(x)=V_{0}(x)-\int_{x}^{1}\left( 4\frac{x^{2}}{y^{3}}-\frac{x}{y^{2}}%
\right) V_{0}(y)dy,  \label{gp17}
\end{equation}%
where the function $V_{0}$ is defined by Eq. (\ref{gp1}) for $n=0$ and $K(p)=
$ $\Delta G(p)$. In the Appendix \ref{app1} it is shown, that the last
relation can be modified to:%
\begin{equation}
V_{-1}(x)=\frac{2}{x}\left( g_{1}(x)+2\int_{x}^{1}\frac{g_{1}(y)}{y}%
)dy\right) .  \label{GP19}
\end{equation}%
Then, in an accordance with Eq. (\ref{gp3}), we obtain 
\begin{equation}
V_{-1}^{\prime }(x)=-\pi M^{3}\Delta G(p);\qquad x=\frac{2p}{M},
\label{gp20}
\end{equation}%
so the last two relations imply%
\begin{equation}
\Delta G(p)=-\frac{2}{\pi M^{3}}\left[ \frac{1}{x}\left(
g_{1}(x)+2\int_{x}^{1}\frac{g_{1}(y)}{y})dy\right) \right] ^{\prime };\qquad
x=\frac{2p}{M},  \label{gp20a}
\end{equation}%
or 
\begin{equation}
\Delta G(p)=\frac{2}{\pi M^{3}x^{2}}\left( 3g_{1}(x)+2\int_{x}^{1}\frac{%
g_{1}(y)}{y}dy-xg_{1}^{\prime }(x)\right) ;\qquad x=\frac{2p}{M}.
\label{gp21}
\end{equation}%
Now we substitute%
\begin{equation}
\Delta q(x)=g_{1}(x)/2,\qquad g_{1}(x)+g_{2}(x)=\int_{x}^{1}\frac{g_{1}(y)}{y%
})dy=\Delta q_{T}(x)/2  \label{gp21b}
\end{equation}%
and in the next we shall consider the flavors separately. The second
equality represents Wanzura-Wilczek relation for twist-2 approximation of $%
g_{2}$, which is valid for present approach as proved in \cite{zav2}. Now
Eq. (\ref{gp21}) in terms of the quark distributions reads%
\begin{equation}
\Delta G_{q}(p)=\frac{1}{\pi M^{3}x^{2}}\left( 3\Delta q(x)+2\int_{x}^{1}%
\frac{\Delta q(y)}{y}dy-x\Delta q^{\prime }(x)\right) ;\qquad x=\frac{2p}{M}
\label{gp21a}
\end{equation}%
or equivalently, with the use of Eqs. (\ref{gp20a}) and (\ref{gp21b}): 
\begin{equation}
\Delta G_{q}(p)=-\frac{1}{\pi M^{3}}\left( \frac{\Delta q(x)+2\Delta q_{T}(x)%
}{x}\right) ^{\prime };\qquad x=\frac{2p}{M}.  \label{gp21c}
\end{equation}%
Obviously the distribution $\Delta G_{q}$ together with the distribution (%
\ref{gp9}) allow to obtain the polarized distributions $G_{q}^{\pm }$ as%
\begin{equation}
G_{q}^{\pm }(p)=\frac{1}{2}\left( G_{q}(p)\pm \Delta G_{q}(p)\right) .
\label{gp22}
\end{equation}%
Distributions $\Delta G_{q}$ and $G_{q}^{\pm }$ measure number of quarks in
the element $d^{3}p$. They can be, similarly as distribution $G_{q}$ in Eq. (%
\ref{gp9a}), replaced by the distributions $\Delta P_{q}$ and $P_{q}^{\pm }$
measuring number of quarks in the element $dp/M$:%
\begin{equation}
\Delta P_{q}(p)=3\Delta q(x)+2\int_{x}^{1}\frac{\Delta q(y)}{y}dy-x\Delta
q^{\prime }(x);\qquad x=\frac{2p}{M},  \label{gp22b}
\end{equation}%
\begin{equation}
P_{q}^{\pm }(p)=\frac{1}{2}\left[ q(x)-xq^{\prime }(x)\pm \left( 3\Delta
q(x)+2\int_{x}^{1}\frac{\Delta q(y)}{y}dy-x\Delta q^{\prime }(x)\right) %
\right] ;\qquad x=\frac{2p}{M}.  \label{gp22d}
\end{equation}%
Obviously the probability distributions should satisfy%
\begin{equation}
\left\vert \Delta G_{q}(p)\right\vert \leq G_{q}(p),  \label{gp22a}
\end{equation}%
which after inserting from Eqs. (\ref{gp21c}) and (\ref{gp9}) implies%
\begin{equation}
\left\vert \left( \frac{\Delta q(x)+2\Delta q_{T}(x)}{x}\right) ^{\prime
}\right\vert \leq -\left( \frac{q(x)}{x}\right) ^{\prime },  \label{gp22c}
\end{equation}%
where positivity of right hand side was required in relation (\ref{gp9b}).
Another self-consistency test of the approach is represented by the
inequality%
\begin{equation}
\left\vert \Delta q(x)\right\vert \leq q(x),  \label{GP30}
\end{equation}%
which is proved\ in the Appendix \ref{app1a}.

\begin{figure}[tbp]
\includegraphics[width=18cm]{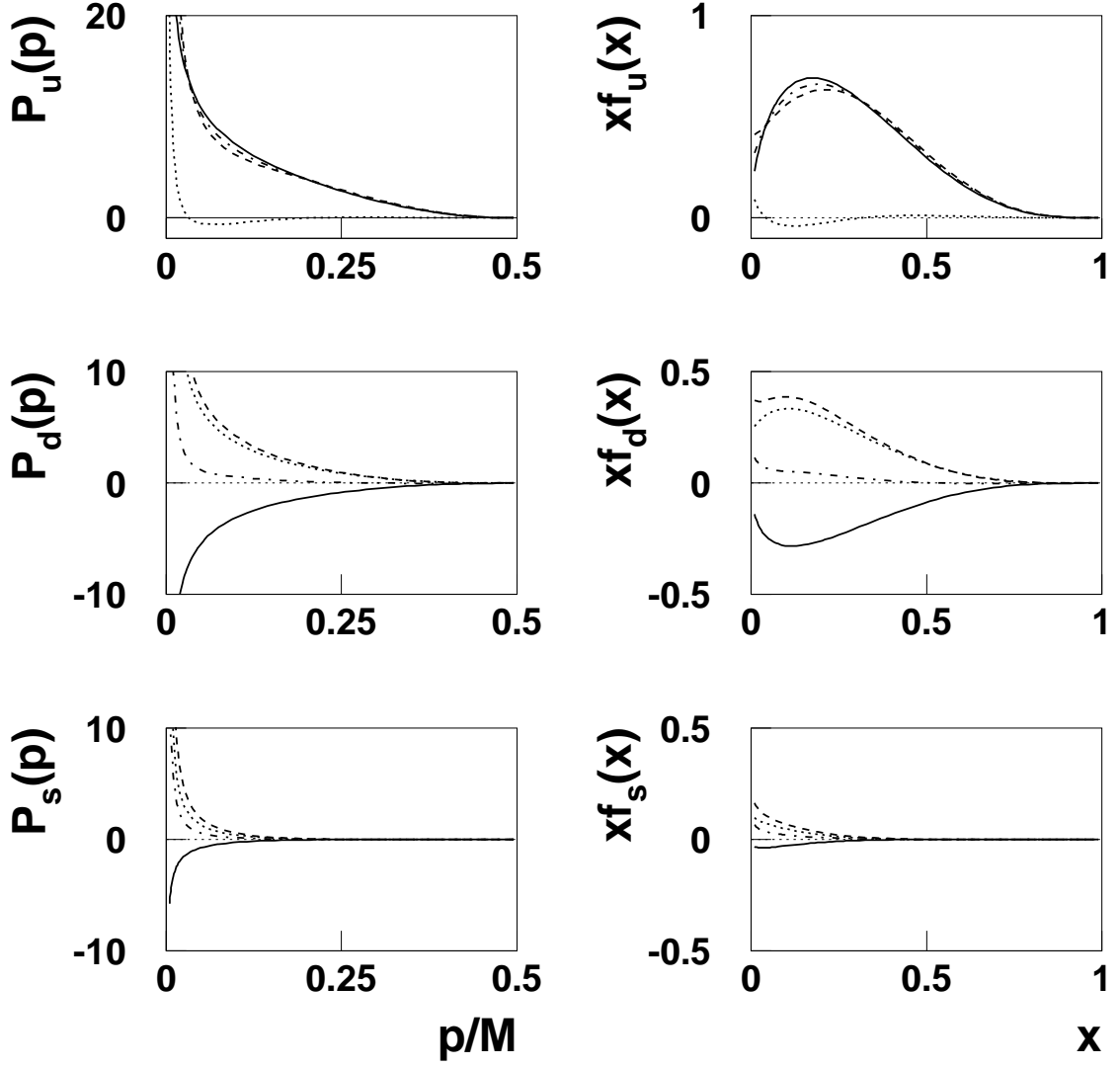}
\caption{Probability distributions $\Delta P_{q},P_{q},P_{q}^{+},P_{q}^{-}$
of $u,d,s$ quarks (left) and related structure functions $\Delta
f_{q},f_{q},f_{q}^{+},$ $f_{q}^{-}$ \ (right) are represented by the solid,
dashed, dash-and-dot and dotted lines.}
\label{fgr2}
\end{figure}

With the use of \ the relation (\ref{sp11e}) one can formally calculate the
partial structure functions corresponding to the subsets of positively and
negatively polarized quarks:%
\begin{equation}
f_{q}^{\pm }(x)=Mx\int G_{q}^{\pm }(p)\delta \left( \frac{p_{0}+p_{1}}{M}%
-x\right) \frac{d^{3}p}{p_{0}}.  \label{gp25bAD}
\end{equation}%
Apparently it holds%
\begin{equation}
f_{q}(x)\equiv f_{q}^{+}(x)+f_{q}^{-}(x)=q(x)  \label{gad1}
\end{equation}%
and one can define also%
\begin{equation}
\Delta f_{q}(x)=f_{q}^{+}(x)-f_{q}^{-}(x)  \label{gad2}
\end{equation}%
or equivalently%
\begin{equation}
\Delta f_{q}(x)=Mx\int \Delta G_{q}(p)\delta \left( \frac{p_{0}+p_{1}}{M}%
-x\right) \frac{d^{3}p}{p_{0}}.  \label{gp23}
\end{equation}%
Obviously%
\begin{equation}
f_{q}^{\pm }(x)=\frac{1}{2}\left( f_{q}(x)\pm \Delta f_{q}(x)\right)
\label{gp23a}
\end{equation}%
and Eq. (\ref{gp22a}) implies%
\begin{equation}
\left\vert \Delta f_{q}(x)\right\vert \leq q(x).  \label{gp23b}
\end{equation}%
Let us note, $f_{q}^{+}+f_{q}^{-}=q$, but $f_{q}^{+}-f_{q}^{-}\neq \Delta q$
in the sense of the relations (\ref{sp11e}) and (\ref{sp11d}). The last
inequality is replaced by equality only in the limit of IMF approach. The
relation (\ref{gp23}) can be written as%
\begin{equation}
\Delta f_{q}(x)=xV_{q,-1}(x),  \label{gp24}
\end{equation}%
where%
\begin{equation}
V_{q,-1}(x)=M\int \Delta G_{q}(p)\delta \left( \frac{p_{0}+p_{1}}{M}%
-x\right) \frac{d^{3}p}{p_{0}}.  \label{24a}
\end{equation}%
At the same time Eq.(\ref{GP19}) can be replaced by%
\begin{equation}
V_{q,-1}(x)=\frac{1}{x}\left( \Delta q(x)+2\int_{x}^{1}\frac{\Delta q(y)}{y}%
)dy\right) .  \label{gp25a}
\end{equation}%
which after inserting from Eq. (\ref{gp24}) gives%
\begin{equation}
\Delta f_{q}(x)=\Delta q(x)+2\int_{x}^{1}\frac{\Delta q(y)}{y})dy.
\label{gp25}
\end{equation}%
This equality together with the relation (\ref{gp23b}) give%
\begin{equation}
\left\vert \Delta q(x)+2\int_{x}^{1}\frac{\Delta q(y)}{y})dy\right\vert \leq
q(x)  \label{gp25b}
\end{equation}%
or equivalently%
\begin{equation}
\left\vert \Delta q(x)+2\Delta q_{T}(x)\right\vert \leq q(x).  \label{gp25c}
\end{equation}%
Now, using the input on the $q(x)$ \cite{pdf} and $\Delta q(x)$ \cite{lss}
(LO at the scale $4GeV^{2}$) one can calculate the distributions $\Delta
P_{q},P_{q},P_{q}^{\pm }$ and related structure functions $\Delta
f_{q},f_{q} $ and $f_{q}^{\pm }$. The result is displayed in Fig. \ref{fgr2}
and one can observe:

\textit{i)} Positivity of distributions $P_{q}^{\pm }$ and $f_{q}^{\pm }$
implies, that self-consistency tests (\ref{gp22a}), (\ref{gp23b}) and their
equivalents (\ref{gp22c}), (\ref{gp25b}) are satisfied with exception of a
small negative disturbance in $G_{u}^{-}(P_{u}^{-})$ and $f_{u}^{-}$. The
possible reason is that the results of the two different procedures for
fitting $q(x)$ and $\Delta q(x)$ are combined and some uncertainty is
unavoidable.

\textit{ii)} The mean value of the distribution $\Delta G_{q}$ can be
estimated as%
\begin{equation}
\left\langle p\right\rangle _{q}=\frac{\int p\Delta G_{q}(p)d^{3}p}{\int
\Delta G_{q}(p)d^{3}p}=\frac{M}{2}\frac{\int_{0}^{1}x\Delta q(x)dx}{%
\int_{0}^{1}\Delta q(x)dx}.  \label{GP26}
\end{equation}%
The proof of this relation is done in the Appendix \ref{app2}. The numerical
calculation gives $0.090GeV/c$ for $u$ and $0.070GeV/c$ for $d$ quarks.
These numbers are well comparable with those calculated from Eq. (\ref{gp16b}%
), which corresponds to the valence quarks. Also the shape of distributions $%
x\Delta f_{u}(x)$ and $x\Delta f_{d}(x)$ is very similar to that of the
valence terms. In other words, results confirm that spin contribution of
quarks comes dominantly from the valence region.

\textit{iii)} Due to input values $\Delta u(x)>0$ and $\Delta d(x)<0$ one
can expect that $P_{u}^{+}\geq P_{u}^{-},$ $P_{d}^{-}\geq P_{d}^{+},$ $%
f_{u}^{+}\geq f_{u}^{-}$ and $f_{d}^{-}\geq f_{d}^{+}$. Besides, the curves
in the figure show, that $P_{u}^{-},P_{d}^{+},f_{u}^{-}$ and $f_{d}^{+}$ are
close to zero, at least in the valence region.

\section{Intrinsic quark motion and orbital momentum}

\label{sec3}The rule of quantum mechanics says, that angular momentum
consists of the orbital and spin part \ $\mathbf{j=l+s}$\ and that in the
relativistic case the $\mathbf{l}$ and $\mathbf{s}$\ are not conserved
separately, but only the total angular momentum $j$\ is conserved.\textit{\ }%
This simple fact was in the context of quarks inside the nucleon pointed out
in \cite{liang}. It means, that only $j^{2},j_{z}$ are well-defined quantum
numbers and corresponding states of the particle with spin $1/2$ are
represented by the bispinor spherical waves \cite{lali}%
\begin{equation}
\psi _{kjlj_{z}}\left( \mathbf{p}\right) =\frac{\delta (p-k)}{p\sqrt{2p_{0}}}%
\left( 
\begin{array}{c}
i^{-l}\sqrt{p_{0}+m}\Omega _{jlj_{z}}\left( \mathbf{\omega }\right) \\ 
i^{-\lambda }\sqrt{p_{0}-m}\Omega _{j\lambda j_{z}}\left( \mathbf{\omega }%
\right)%
\end{array}%
\right) ,  \label{sp19}
\end{equation}%
where $\mathbf{\omega }=\mathbf{p}/p,$\ $l=j\pm \frac{1}{2},\ \lambda =2j-l$
($l$ defines the parity) and 
\begin{eqnarray*}
\Omega _{j,l,j_{z}}\left( \mathbf{\omega }\right) &=&\left( 
\begin{array}{c}
\sqrt{\frac{j+j_{z}}{2j}}Y_{l,j_{z}-1/2}\left( \mathbf{\omega }\right) \\ 
\sqrt{\frac{j-j_{z}}{2j}}Y_{l,j_{z}+1/2}\left( \mathbf{\omega }\right)%
\end{array}%
\right) ;\quad l=j-\frac{1}{2}, \\
\Omega _{j,l,j_{z}}\left( \mathbf{\omega }\right) &=&\left( 
\begin{array}{c}
-\sqrt{\frac{j-j_{z}+1}{2j+2}}Y_{l,j_{z}-1/2}\left( \mathbf{\omega }\right)
\\ 
\sqrt{\frac{j+j_{z}+1}{2j+2}}Y_{l,j_{z}+1/2}\left( \mathbf{\omega }\right)%
\end{array}%
\right) ;\quad l=j+\frac{1}{2}.
\end{eqnarray*}%
States are normalized as:%
\begin{equation}
\int \psi _{k^{\prime }j^{\prime }l^{\prime }j_{z}^{\prime }}^{\dagger
}\left( \mathbf{p}\right) \psi _{kjlj_{z}}\left( \mathbf{p}\right)
d^{3}p=\delta (k-k^{\prime })\delta _{jj^{\prime }}\delta _{ll^{\prime
}}\delta _{j_{z}j_{z}^{\prime }}.  \label{sp20}
\end{equation}%
The wavefunction (\ref{sp19}) is simplified for $j=j_{z}=1/2$ and $l=0$.
Taking into account that%
\[
Y_{00}=\frac{1}{\sqrt{4\pi }},\qquad Y_{10}=i\sqrt{\frac{3}{4\pi }}\cos
\theta ,\qquad Y_{11}=-i\sqrt{\frac{3}{8\pi }}\sin \theta \exp \left(
i\varphi \right) , 
\]%
one gets:%
\begin{equation}
\psi _{kjlj_{z}}\left( \mathbf{p}\right) =\frac{\delta (p-k)}{p\sqrt{8\pi
p_{0}}}\left( 
\begin{array}{c}
\sqrt{p_{0}+m}\left( 
\begin{array}{c}
1 \\ 
0%
\end{array}%
\right) \\ 
-\sqrt{p_{0}-m}\left( 
\begin{array}{c}
\cos \theta \\ 
\sin \theta \exp \left( i\varphi \right)%
\end{array}%
\right)%
\end{array}%
\right) .  \label{sp22}
\end{equation}%
Let us note, that $j=1/2$ is the minimum angular momentum for the particle
with spin $1/2.$ If one consider the quark state as a superposition%
\begin{equation}
\Psi \left( \mathbf{p}\right) =\int a_{k}\psi _{kjlj_{z}}\left( \mathbf{p}%
\right) dk;\quad \int a_{k}^{\star }a_{k}dk=1  \label{sp21}
\end{equation}%
then its average spin contribution to the total angular momentum reads:%
\begin{equation}
\left\langle s\right\rangle =\int \Psi ^{\dagger }\left( \mathbf{p}\right)
\Sigma _{z}\Psi \left( \mathbf{p}\right) d^{3}p;\qquad \Sigma _{z}=\frac{1}{2%
}\left( 
\begin{array}{cc}
\sigma _{z} & \cdot \\ 
\cdot & \sigma _{z}%
\end{array}%
\right) .  \label{sp26}
\end{equation}%
After inserting from Eqs. (\ref{sp22}), (\ref{sp21}) into (\ref{sp26}) one
gets

\begin{equation}
\left\langle s\right\rangle =\int a_{p}^{\star }a_{p}\frac{\left(
p_{0}+m\right) +\left( p_{0}-m\right) \left( \cos ^{2}\theta -\sin
^{2}\theta \right) }{16\pi p^{2}p_{0}}d^{3}p=\frac{1}{2}\int a_{p}^{\star
}a_{p}\left( \frac{1}{3}+\frac{2m}{3p_{0}}\right) dp.  \label{sp24}
\end{equation}%
Since\ $j=1/2$, the last relation implies for the quark orbital momentum:%
\begin{equation}
\left\langle l\right\rangle =\frac{1}{3}\int a_{p}^{\star }a_{p}\left( 1-%
\frac{m}{p_{0}}\right) dp.  \label{sp25}
\end{equation}%
It means that for quarks in the state $j=j_{z}=1/2$ there are the extreme
scenarios:

\textit{i)} Massive and static quarks ($p_{0}=m$), which implies $%
\left\langle s\right\rangle =j=1/2$ and $\left\langle l\right\rangle =0$.\
This is evident, since without kinetic energy no orbital momentum can be
generated.

\textit{ii)}\ Massless quarks $\left( m\ll p_{0}\right) $, which implies $%
\left\langle s\right\rangle =1/6$ and $\left\langle l\right\rangle =1/3$.

Generally, for $p_{0}\geq m$, one gets $1/3\leq \left\langle s\right\rangle
/j\leq 1.$ In other words, for the states with $p_{0}>m$ part of the total
angular momentum $j=1/2$ is necessarily generated by the orbital momentum.
This is a consequence of quantum mechanics, and not a consequence of
particular model. If one assumes the quark effective mass of the order
thousandths and intrinsic momentum of the order of tenth of $GeV$, which is
quite realistic consideration, then the second scenario is clearly
preferred. Further, the mean kinetic energy corresponding to the
superposition (\ref{sp21}) reads%
\begin{equation}
\left\langle E_{kin}\right\rangle =\int a_{p}^{\star }a_{p}E_{kin}dp;\qquad
E_{kin}=p_{0}-m  \label{gp30}
\end{equation}%
and at the same time the Eq. (\ref{sp25}) can be rewritten as%
\begin{equation}
\left\langle l\right\rangle =\frac{1}{3}\int a_{p}^{\star }a_{p}\frac{E_{kin}%
}{p_{0}}dp.  \label{gp31}
\end{equation}%
It is evident, that for fixed $j=1/2$ both the quantities are in the nucleon
rest frame almost equivalent: more kinetic energy generates more orbital
momentum and vice versa.

Further, the average spin part $\left\langle s\right\rangle $ of the total
angular momentum $j=1/2$ related to single quark according to Eq. (\ref{sp24}%
) can be compared to the integral%
\begin{equation}
\Gamma _{1}=\int_{0}^{1}g_{1}(x)dx,  \label{sp4}
\end{equation}%
which measures total quark spin contribution to the nucleon spin. For the $%
g_{1}$ from Eq. (\ref{sp10}) this integral reads%
\begin{equation}
\Gamma _{1}=\frac{1}{2}\int \Delta G(p_{0})\left( \frac{1}{3}+\frac{2m}{%
3p_{0}}\right) d^{3}p.  \label{sp17}
\end{equation}%
Dependence of both \ the integrals (\ref{sp24}) and (\ref{sp17}) on
intrinsic motion is controlled by the same term $\left( 1/3+2m/3p_{0}\right) 
$, which in both the cases has origin in the covariant kinematics of the
particle with $s=1/2$. In fact, the \ procedures for calculation of these
integrals are based on the two different representations of the solutions of
\ Dirac equation: the plane waves (\ref{sp8}) and spherical waves (\ref{sp22}%
). It is apparent that for the scenario of massless quarks $\left( m\ll
p_{0}\right) $, due to necessary presence of the orbital motion, both the
integrals $\Gamma _{1}$ and $\left\langle s\right\rangle $ will be roughly
three times less, than for the scenario of massive and static quarks $\left(
m\simeq p_{0}\right) $. What is the underlying physics behind the interplay
between the spin and orbital momentum? Actually, speaking about the spin of
the particle represented by the state (\ref{sp8}),\ one should take into
account:

\textit{a)} Definite projection of the spin in the direction $\mathbf{n}$ is
well-defined quantum number only for the particle at rest ($p=0$) or for the
particle moving in the the direction $\mathbf{n}$, i.e. $\mathbf{p/}p\mathbf{%
=\pm n}$. In these cases we have%
\begin{equation}
s=u^{\dagger }\left( \mathbf{p},\lambda \mathbf{n}\right) \mathbf{n\Sigma }%
u\left( \mathbf{p},\lambda \mathbf{n}\right) =\pm 1/2.  \label{gp32}
\end{equation}

\textit{b)} In other cases, as shown in the Appendix \ref{app3}, only
inequality%
\begin{equation}
\left\langle s\right\rangle =\left\vert u^{\dagger }\left( \mathbf{p}%
,\lambda \mathbf{n}\right) \mathbf{n\Sigma }u\left( \mathbf{p},\lambda 
\mathbf{n}\right) \right\vert <1/2  \label{GP33}
\end{equation}%
is satisfied. Roughly speaking, the result of measuring the spin (of a
quark) depends on its momentum in the defined reference frame (nucleon rest
frame). This obvious effect acts also in the states, which are represented
by the superposition of the plane waves (\ref{sp8}) with different momenta $%
\mathbf{p}$ and resulting in $\left\langle \mathbf{p}\right\rangle =0$, but $%
\left\langle \mathbf{p}^{2}\right\rangle >0$. In \cite{zav1} we showed, that
averaging of the spin projection (\ref{GP33}) over the spherical momentum
distribution gives the result equivalent to (\ref{sp17}). The state (\ref%
{sp21}) can be also decomposed into plane waves having spherical momentum
distribution and the spin mean value given by Eq. (\ref{sp24}). Well-defined
quantum numbers $j=j_{z}=1/2$ imply, that the spin reduction due to
increasing intrinsic kinetic energy \ is compensated by the increasing
orbital momentum.

Now, what the preferred scenario of massless quarks ($\left\langle
m/p_{0}\right\rangle \ll 1$) implies for the spin structure of whole
nucleon, what are the integral quark spin and orbital contributions to the
nucleon spin? Obviously using some input on the total quark longitudinal
polarization $\Delta \Sigma ,$ one can estimate the relative quark spin and
orbital contributions as:%
\begin{equation}
S=\Delta \Sigma ,\qquad L=2\Delta \Sigma ;\qquad \Delta \Sigma
=\sum_{q}\int_{0}^{1}\Delta q(x)dx.  \label{gp34}
\end{equation}%
At the same time our approach can be compared with the calculation based on
the chiral quark soliton model (CQSM) \cite{waka},\cite{waka1}, where
significant role of the quark orbital momentum is considered as well. In
Tab. \ref{tb2} some results of both models are shown. However, in spite of
some similarity between the two sets of numbers, there are substantial
differences between both the approaches. Let us mention at least the two,
which seem to be most evident:

\textit{1)} Presence of significant fraction of the orbital momentum in the
CQSM apparently follows from dynamics inherent in the model. On the other
hand, in our approach the important role of the orbital momentum follows
from kinematics, so it should not be too sensitive to details of inherent
dynamics. Actually effect takes place in LO when quarks interacting with
probing photon can be effectively described as free fermions in the states
like (\ref{sp21}) with sufficiently low effective ratio $\left\langle
m/p_{0}\right\rangle $ which controls the fraction of orbital momentum (\ref%
{sp25}). Of course, value of this ratio itself is question of the dynamics.

\textit{2)} In the CQSM antiquarks are predicted to have opposite signs of
the spin and orbital contributions. In our approach both contributions are
proportional and have the same signs regardless of flavor or antiflavor.

Last comment concerns the total quark angular momentum $\ J,$ by which a
room for the gluon contribution $J_{g}$ is\ defined. Results in Tab. \ref%
{tb2} related to the CQSM suggest, that higher $Q^{2}$\ implies greater
gluon contribution. Our results suggest, that gluon contribution can be
rather sensitive to the longitudinal polarization: for $\Delta \Sigma \simeq
1/3,$ $0.3$ and $0.2$ the gluon contribution can represent  $\simeq 0,$ $10$
and $40\%$ respectively. Let us remark, that the value empirically known 
\cite{waka1}   
\begin{equation}
\Delta \Sigma \simeq 0.2-0.35  \label{gp35}
\end{equation}%
does not exclude any of these possibilities.

\section{Summary and conclusion}

We studied covariant version of the QPM with spherically symmetric
distributions of \ the quark momentum in the nucleon rest frame. The main
results obtained in this paper can be summarized as follows.

1) Relations between the distribution functions $q(x),\Delta q(x)$ and
corresponding 3D quark momentum distributions $G_{q}^{\pm }(p)=G_{q}(p)\pm
\Delta G_{q}(p)$ are obtained. In this way the momentum distributions of
positively and negatively polarized quarks $G_{q}^{\pm }(p)$ are calculated
from the input, which is obtained from experimentally measured structure
functions $F_{2}$ and $g_{1}$. At the same time these relations, due to
positivity of probabilistic distributions $G_{q}$ and $G_{q}^{\pm }$, imply
some inequalities for $q(x),\Delta q(x)$. We proved, that these constraints,
serving as self-consistency tests of the approach, are satisfied.

2) We showed, that important role of the quark orbital momentum emerges as a
direct consequence of a covariant description. Since in relativistic case
only the total angular momentum \ $\mathbf{j=l+s}$\ is well-defined quantum
number, there arises some interplay between its spin and orbital parts. For
the quark in the state with definite projection $j_{z}=1/2$ in the nucleon
rest frame, as a result of this interplay, its spin part is reduced in favor
of the orbital one. The role of orbital motion increases with the rate of
quark intrinsic motion; for $\left\langle m/p_{0}\right\rangle \ll 1$\ its
fraction reaches $\left\langle l_{z}\right\rangle =2/6$ whereas $%
\left\langle s_{z}\right\rangle =1/6$ only. Simultaneously this effect is
truly reproduced also in the formalism of\ structure functions and in this
connection some implications for the global nucleon spin structure were
suggested. \bigskip

\textbf{Acknowledgments}

\textit{This work was supported by the project AV0Z10100502 of the Academy
of Sciences of the Czech Republic. I am grateful to Anatoli Efremov and Oleg
Teryaev for many useful discussions and valuable comments.}

\appendix

\section{Structure functions in the approach of infinite momentum frame}

\label{app0}The necessary condition for obtaining equalities (\ref{sp11a}) -
(\ref{sp11b}) is the covariant relation%
\begin{equation}
p_{\alpha }=yP_{\alpha },  \label{b2}
\end{equation}%
which implies%
\begin{equation}
m=yM  \label{b3}
\end{equation}%
and $\mathbf{p}=0$ in the nucleon rest frame and $p_{T}=0$ in the IMF.

For calculation of the integrals (\ref{sp9c}) and (\ref{sp9d}) in the IMF
approach one can substitute $p$ by $yP$ and $d^{3}p/p_{0}$ by $\pi
dp_{T}^{2}dy/y$. Then, after some algebra the structure functions (\ref{sp9b}%
) read%
\begin{equation}
F_{1}(x)=\frac{1}{2}Mx\int G\left( yM\right) \delta \left( y-x\right) \pi
dp_{T}^{2}\frac{dy}{y},\qquad F_{2}(x)=Mx^{2}\int G\left( yM\right) \delta
\left( y-x\right) \pi dp_{T}^{2}\frac{dy}{y}.  \label{b6}
\end{equation}%
Since the approximation (\ref{b2}) implies sharply peaked distribution at $%
p_{T}^{2}\rightarrow 0$, one can identify%
\begin{equation}
MG_{q}\left( yM\right) \pi dp_{T}^{2}=q(y)  \label{b7}
\end{equation}%
and then the Eqs. (\ref{sp11a}) and (\ref{b6}) after integrating are
equivalent.

In the same way the equalities (\ref{cra31})-(\ref{cr31}) can be modified.
Taking into account that $pS\rightarrow yPS=0$, one obtain%
\begin{equation}
g_{1}(x)=\frac{m}{2}\int \Delta G\left( yM\right) \delta \left( y-x\right)
\pi dp_{T}^{2}\frac{dy}{y},\qquad g_{2}(x)=0.  \label{b8}
\end{equation}%
If we put%
\begin{equation}
M\Delta G_{q}\left( yM\right) \pi dp_{T}^{2}=\Delta q(y)  \label{b9}
\end{equation}%
\ and take into account Eq. (\ref{b3}), then it is obvious, that the Eqs. (%
\ref{sp11b}) and (\ref{b8}) are equivalent.

\section{Proof of the relation ({\protect\ref{GP19}})}

\label{app1}In the paper \cite{zav2} we proved relation%
\begin{equation}
\frac{V_{j}^{\prime }(x)}{V_{k}^{\prime }(x)}=\left( \frac{x}{2}+\frac{%
x_{0}^{2}}{2x}\right) ^{j-k};\qquad x_{0}=\frac{m}{M},  \label{a1}
\end{equation}%
which for $m\rightarrow 0$ implies%
\begin{equation}
V_{0}(x)=\frac{1}{2}\left( xV_{-1}(x)+\int_{0}^{x}V_{-1}(y)dy\right) .
\label{a2}
\end{equation}%
After inserting $V_{0}$ from this relation to Eq. (\ref{gp17}) one gets%
\begin{eqnarray}
g_{1}(x) &=&\frac{1}{2}\left( xV_{-1}(x)+\int_{0}^{x}V_{-1}(y)dy\right)
\label{a3} \\
&&-2x^{2}\left( \int_{x}^{1}\frac{V_{-1}(y)}{y^{2}}dy+\int_{x}^{1}\frac{1}{%
y^{3}}\int_{y}^{1}V_{-1}(z)dzdy\right)  \nonumber \\
&&+\frac{1}{2}x\left( \int_{x}^{1}\frac{V_{-1}(y)}{y}dy+\int_{x}^{1}\frac{1}{%
y^{2}}\int_{y}^{1}V_{-1}(z)dzdy\right) .  \nonumber
\end{eqnarray}%
The double integrals can be reduced by integration by parts with the use of
\ formula%
\begin{equation}
\int_{x}^{1}a(y)\left( \int_{y}^{1}b(z)dz\right) dy=\int_{x}^{1}\left(
A(y)-A(x)\right) b(y)dy;\qquad A^{\prime }(x)=a(x),  \label{a4}
\end{equation}%
then the relation (\ref{a3}) is simplified:%
\begin{equation}
g_{1}(x)=\frac{1}{2}xV_{-1}(x)-x^{2}\int_{x}^{1}\frac{V_{-1}(y)}{y^{2}}dy.
\label{a5}
\end{equation}%
In the next step we extract $V_{-1}$ from this relation. After the
substitution $V(x)=V_{-1}(x)/x$ the relation reads%
\begin{equation}
\frac{g_{1}(x)}{x^{2}}=\frac{1}{2}V(x)-\int_{x}^{1}\frac{V(y)}{y}dy,
\label{a6}
\end{equation}%
which implies the differential equation for $V(x)$:%
\begin{equation}
\frac{1}{2}V^{\prime }(x)+\frac{V(x)}{x}=\left( \frac{g_{1}(x)}{x^{2}}%
\right) ^{\prime }.  \label{a7}
\end{equation}%
The corresponding homogeneous equation%
\begin{equation}
\frac{1}{2}V^{\prime }(x)+\frac{V(x)}{x}=0  \label{a8}
\end{equation}%
gives the solution%
\begin{equation}
V(x)=\frac{C}{x^{2}},  \label{a9}
\end{equation}%
which after inserting to Eq. (\ref{a7}) gives%
\begin{equation}
C^{\prime }(x)=2x^{2}\left( \frac{g_{1}(x)}{x^{2}}\right) ^{\prime }.
\label{a10}
\end{equation}%
After integration one easily gets the relation inverse to Eq. (\ref{a5}):%
\begin{equation}
V_{-1}(x)=\frac{2}{x}\left( g_{1}(x)+2\int_{x}^{1}\frac{g_{1}(y)}{y}%
dy\right) ,  \label{a11}
\end{equation}%
which coincides with Eq. (\ref{GP19}).

\section{Proof of the relation ({\protect\ref{GP30}})}

\label{app1a}Relations (\ref{sp11e}) and (\ref{sp11d}) imply that inequality
(\ref{GP30}) is satisfied if%
\begin{equation}
p_{0}+p_{1}\geq \left\vert m+p_{1}+\frac{p_{1}^{2}}{p_{0}+m}\right\vert
=\left\vert p_{0}+p_{1}-\frac{p_{T}^{2}}{p_{0}+m}\right\vert .  \label{b1}
\end{equation}%
There are two cases:

\textit{a)} $p_{0}+p_{1}-p_{T}^{2}/\left( p_{0}+m\right) \geq 0$, then
instead of the relation (\ref{b1}) one can write 
\begin{equation}
p_{0}+p_{1}\geq p_{0}+p_{1}-\frac{p_{T}^{2}}{p_{0}+m},  \label{b4}
\end{equation}%
which is always satisfied.

\textit{b)} $p_{0}+p_{1}-p_{T}^{2}/\left( p_{0}+m\right) <0$, then the
relation (\ref{b1}) is equivalent to%
\begin{equation}
p_{0}+p_{1}\geq -p_{0}-p_{1}+\frac{p_{T}^{2}}{p_{0}+m}\Leftrightarrow
2\left( p_{0}+p_{1}\right) \geq \frac{p_{T}^{2}}{p_{0}+m}.  \label{b5}
\end{equation}%
Since%
\[
2p_{0}\geq p_{0}-p_{1}\Rightarrow 2\left( p_{0}+m\right) \geq
p_{0}-p_{1}\Rightarrow 2\left( p_{0}+m\right) \left( p_{0}+p_{1}\right) \geq
\left( p_{0}-p_{1}\right) \left( p_{0}+p_{1}\right) 
\]%
\[
\Rightarrow 2\left( p_{0}+m\right) \left( p_{0}+p_{1}\right) \geq
p_{T}^{2}\Rightarrow 2\left( p_{0}+p_{1}\right) \geq \frac{p_{T}^{2}}{p_{0}+m%
}, 
\]%
which means, that (\ref{b5}) is always satisfied. In this way the relations (%
\ref{b1}) and ({\ref{GP30}}) are proved.

\section{Proof of the relation ({\protect\ref{GP26}})}

\label{app2}The relation (\ref{gp21}) implies%
\begin{equation}
\int \Delta G_{q}(p)d^{3}p=\frac{1}{2}\int_{0}^{1}\left( 3\Delta
q(x)+2\int_{x}^{1}\frac{\Delta q(y)}{y})dy-x\Delta q^{\prime }(x)\right) dx
\label{a12}
\end{equation}%
and%
\begin{equation}
\int p\Delta G(p)d^{3}p=\frac{M}{4}\int_{0}^{1}\left( 3x\Delta
q(x)+2x\int_{x}^{1}\frac{\Delta q(y)}{y})dy-x^{2}\Delta q^{\prime
}(x)\right) dx;\qquad x=\frac{2p}{M}.  \label{a13}
\end{equation}%
If one denotes%
\begin{equation}
\Gamma _{1}^{q}=\int_{0}^{1}\Delta q(x)dx,\qquad \Gamma
_{2}^{q}=\int_{0}^{1}x\Delta q(x)dx,  \label{a14}
\end{equation}%
then integration by parts gives%
\begin{equation}
\int_{0}^{1}\int_{x}^{1}\frac{\Delta q(y)}{y})dydx=\Gamma _{1}^{q},\qquad
\int_{0}^{1}x\Delta q^{\prime }(x)dx=-\Gamma _{1}^{q}  \label{a15}
\end{equation}%
and%
\begin{equation}
\int_{0}^{1}2x\int_{x}^{1}\frac{\Delta q(y)}{y})dydx=\Gamma _{2}^{q},\qquad
\int_{0}^{1}x^{2}\Delta q^{\prime }(x)dx=-2\Gamma _{2}^{q}.  \label{a16}
\end{equation}%
Now, one can easily express the ratio%
\begin{equation}
\frac{\int p\Delta G(p)d^{3}p}{\int \Delta G(p)d^{3}p}=\frac{M}{2}\frac{%
\Gamma _{2}^{q}}{\Gamma _{1}^{q}},  \label{a17}
\end{equation}%
in this way the relation ({\ref{GP26}}) is proved.

\section{Proof of the relation ({\protect\ref{GP33}})}

\label{app3}With the use of rule%
\begin{equation}
\mathbf{p\sigma \cdot n\sigma }+\mathbf{n\sigma \cdot p\sigma =2pn}
\label{a18}
\end{equation}%
the term in Eq. ({\ref{GP33}}) can be modified as

\begin{eqnarray}
u^{\dagger }\left( \mathbf{p},\lambda \mathbf{n}\right) \mathbf{n\Sigma }%
u\left( \mathbf{p},\lambda \mathbf{n}\right) &=&\frac{1}{2N}\phi _{\lambda 
\mathbf{n}}^{\dagger }\left( \mathbf{n\sigma +}\frac{\mathbf{p}\mathbf{%
\sigma \cdot n\sigma \cdot p}\mathbf{\sigma }}{\left( p_{0}+m\right) ^{2}}%
\right) \phi _{\lambda \mathbf{n}}  \label{a19} \\
&=&\frac{1}{2N}\phi _{\lambda \mathbf{n}}^{\dagger }\left( \mathbf{n\sigma +}%
\frac{\mathbf{p}\mathbf{\sigma \cdot }\left( -\mathbf{p\sigma \cdot n\sigma
+2pn}\right) }{\left( p_{0}+m\right) ^{2}}\right) \phi _{\lambda \mathbf{n}}
\nonumber \\
&=&\frac{1}{2N}\phi _{\lambda \mathbf{n}}^{\dagger }\left( \mathbf{n\sigma }%
\left( \mathbf{1-}\frac{\mathbf{p}^{2}}{\left( p_{0}+m\right) ^{2}}\right) 
\mathbf{+}\frac{\mathbf{2pn\cdot p}\mathbf{\sigma }}{\left( p_{0}+m\right)
^{2}}\right) \phi _{\lambda \mathbf{n}}  \nonumber \\
&=&\frac{1}{2p_{0}}\phi _{\lambda \mathbf{n}}^{\dagger }\left( m\cdot 
\mathbf{n\sigma +}\frac{\mathbf{pn\cdot p}\mathbf{\sigma }}{p_{0}+m}\right)
\phi _{\lambda \mathbf{n}}.  \nonumber
\end{eqnarray}%
Since%
\begin{equation}
\left| \phi _{\lambda \mathbf{n}}^{\dagger }\mathbf{n\sigma }\phi _{\lambda 
\mathbf{n}}\right| =1,\qquad \left| \phi _{\lambda \mathbf{n}}^{\dagger }%
\mathbf{p\sigma }\phi _{\lambda \mathbf{n}}\right| \leq p,\qquad \mathbf{pn=}%
p\mathbf{\cos \alpha ,}  \label{a20}
\end{equation}%
it follows%
\begin{equation}
\left| u^{\dagger }\left( \mathbf{p},\lambda \mathbf{n}\right) \mathbf{%
n\Sigma }u\left( \mathbf{p},\lambda \mathbf{n}\right) \right| \leq \frac{1}{%
2p_{0}}\left( m\mathbf{+}\frac{p^{2}}{p_{0}+m}\right) =\frac{1}{2}.
\label{a21}
\end{equation}%
Obviously 
\begin{equation}
\left| u^{\dagger }\left( \mathbf{p},\lambda \mathbf{n}\right) \mathbf{%
n\Sigma }u\left( \mathbf{p},\lambda \mathbf{n}\right) \right| =\frac{1}{2}
\label{a22}
\end{equation}%
only for $\mathbf{p/}p\mathbf{=\pm n}$ or $p=0\mathbf{.}$

\newpage 

\begin{table}[tbp]
\includegraphics[width=18cm]{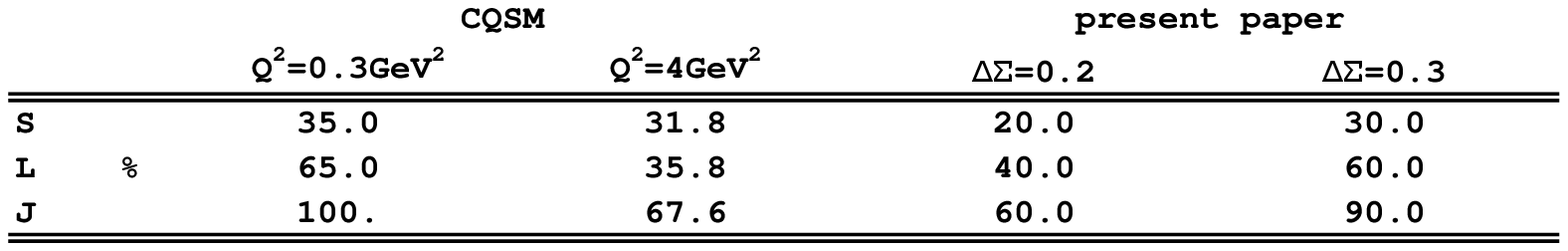}
\caption{Relative integral contributions of the quark spins $(S),$ orbital
momenta $(L)$ and their sum $(J)$\ to the total nucleon spin. Results of our
calculation (right) and prediction of the CQSM model (left).}
\label{tb2}
\end{table}

\end{document}